\documentclass{article}

\usepackage{graphicx} % Required for inserting images
\usepackage{siunitx}
\usepackage{subfigure}
\usepackage{bm}
\usepackage{moreverb}
\usepackage{diagbox}
\usepackage{multirow}
\usepackage{authblk}
\usepackage{amsmath}
\usepackage{pdfpages}
\usepackage{geometry}
\usepackage{lineno}
\usepackage[title]{appendix}

\usepackage{algorithm}
\usepackage{algpseudocode}

\usepackage[driverfallback=dvipdfm,colorlinks,bookmarksopen,bookmarksnumbered,citecolor=red,urlcolor=red]{hyperref}

\usepackage{tikz}
\usetikzlibrary{
  calc,%
  trees,%
  positioning,%
  arrows,%
  chains,%
  shapes.geometric,%
  decorations.pathreplacing,%
 decorations.pathmorphing,%
  shapes,%
  matrix,%
 shapes.symbols,%
 intersections,%
  fit,%
  patterns,%
  angles,%
  quotes%
 }

\tikzstyle{startstop} = [rectangle, rounded corners, minimum width=3cm, minimum height=1cm,text centered, draw=black]
\tikzstyle{io} = [trapezium, trapezium left angle=70, trapezium right angle=110, minimum width=3cm, minimum height=1cm, text centered, draw=black]
\tikzstyle{arrow} = [thick,->,>=stealth]
\tikzstyle{decision} = [diamond, draw,minimum width=2.5cm,minimum height=1.5cm text width=4.5em, text centered, inner sep=0pt]

\definecolor{color1}{HTML}{0000AD} % blau
\definecolor{color2}{HTML}{FF4500} % rot
\definecolor{color3}{HTML}{FFA500} % gelb
\definecolor{color4}{HTML}{00BB00} % gruen
\definecolor{color5}{HTML}{9400D3} % lila
\definecolor{color6}{HTML}{800000} % bordeauxrot/braun
\definecolor{color7}{HTML}{000000} % schwarz
\definecolor{color8}{HTML}{0000FF} % blau heller
\definecolor{color9}{HTML}{FF0000} % rot heller
\definecolor{color10}{HTML}{11CCEE} % türkis heller
\definecolor{color11}{HTML}{606060} % grau

\tikzset{%
    body/.style={inner sep=0pt,outer sep=0pt,shape=rectangle,draw,thick,pattern=north east lines wide},
    dimen/.style={<->,>=latex,thin,every rectangle node/.style={fill=white,midway}},
    symmetry/.style={dashed,thin},
}

\tikzstyle{line1} = [color=color7,semithick]
\tikzstyle{line2} = [color=color2,densely dotted,thick]
\tikzstyle{line3} = [color=color1,densely dashed,thick]
\tikzstyle{line4} = [color=color4,dash dot,thick]
\tikzstyle{line5} = [color=color5,dash dot dot,thick]
\tikzstyle{line6} = [color=color6,densely dotted,thick]
\tikzstyle{line7} = [color=color10,densely dashed,thick]
\tikzstyle{line8} = [color=color3,thick]

\tikzstyle{line10} = [color=color9,densely dotted,thick]% rot heller
\tikzstyle{line11} = [color=color12,dash dot,thick]% dunkelgr�n
\tikzstyle{line12} = [color=color13,dash dot,thick]% hellgr�n

\tikzstyle{line13} = [color=color4,dash dot,line width=1pt]% gruen
\tikzstyle{line14} = [color=color8,densely dashed,line width=1pt]% blau

\tikzstyle{line15} = [color=color1,densely dashed,thick]% blau
\tikzstyle{line16} = [color=color12,dash dot,thick]% dunkelgruen
\tikzstyle{line17} = [color=color6,densely dotted,thick]% dunkelrot

\tikzstyle{mark1} = [color=color7,mark=x,mark size=2pt,mark options=solid,semithick]
\tikzstyle{mark2} = [color=color2,mark=square,mark size=2pt,mark options=solid,semithick]
\tikzstyle{mark3} = [color=color1,mark=triangle,mark size=2pt,mark options=solid,semithick]
\tikzstyle{mark4} = [color=color4,mark=o,mark size=2pt,mark options=solid,semithick]
\tikzstyle{mark5} = [color=color2,mark=square*,mark size=2pt,mark options=solid,semithick]
\tikzstyle{mark6} = [color=color1,mark=triangle*,mark size=2pt,mark options=solid,semithick]
\tikzstyle{mark7} = [color=color4,mark=*,mark size=2pt,mark options=solid,semithick]
\tikzstyle{mark8} = [color=color2,mark=x,mark size=2.5pt,mark options=solid,semithick]
\tikzstyle{mark9} = [color=color1,mark=diamond,mark size=2pt,mark options=solid,semithick]
\tikzstyle{mark10} = [color=color4,mark=asterisk,mark size=2.5pt,mark options=solid,semithick]
\tikzstyle{mark11} = [color=color7,mark=*,mark size=2pt,mark options=solid,semithick]
\tikzstyle{mark12} = [color=color5,mark=*,mark size=2pt,mark options=solid,semithick]
\tikzstyle{mark13} = [color=color7,mark=*,mark size=1pt,mark options=solid,semithick]
\tikzstyle{mark14} = [color=color3,mark=x,mark size=2.5pt,mark options=solid,semithick]

\tikzstyle{mark15} = [color=color7,mark=*,mark size=0.75pt,mark options=solid,semithick]
\tikzstyle{mark16} = [color=color6,mark=*,mark size=1pt,mark options=solid,semithick]
\tikzstyle{mark17} = [color=color8,mark=x,mark size=2.5pt,mark options=solid,semithick]
\tikzstyle{mark18} = [color=color8,mark=x,mark size=4.0pt,mark options=solid,semithick]

\newcommand\BibTeX{{\rmfamily B\kern-.05em \textsc{i\kern-.025em b}\kern-.08em
T\kern-.1667em\lower.7ex\hbox{E}\kern-.125emX}}

\begin{document}

\title{A two-phase Volume of Fluid approach to model rigid-perfectly plastic granular materials}
\author[1]{Wibke D\"usterh\"oft-Wriggers}
\author[1]{Svenja Schubert}
\author[1]{Thomas Rung}
\affil[1]{Hamburg University of Technology, Inst. Fluid Dynamics and Ship Theory,  Am Schwarzenberg-Campus 4, D-21073 Hamburg, Germany}
\affil[]{corresponding author: wibke.wriggers@tuhh.de}

\date{}

\renewcommand\Affilfont{\itshape\small}
\maketitle

%\linenumbers

\begin{abstract}
Granular flow problems characterized by large deformations are widespread in various applications, including coastal and geotechnical engineering.
The paper deals with the application of a rigid-perfectly plastic two-phase model extended by the Drucker-Prager yield criterion to simulate granular media with a finite volume flow solver (FV). The model refers to the combination of a Bingham fluid and an Eulerian strain measure to assess the failure region of granular dam slides. A monolithic volume-of-fluid (VoF) method is used to distinguish between the air and granular phases, both governed by the incompressible Navier-Stokes equations.
The numerical framework enables modeling of large displacements and arbitrary shapes for large-scale applications.  The displayed validation and verification focuses on the rigid-perfectly plastic material model for non-cohesive and cohesive materials with varying angles of repose. Results indicate a good agreement of the predicted soil surface and strain results with experimental and numerical data.

\end{abstract}


\vspace{-6pt}

\section{Introduction}
\vspace{-2pt}
Granular flow problems characterized by large deformations are a concern in various domains, encompassing coastal/ocean engineering and geotechnical engineering. Applications of the model are, e.g., embankment failures, debris flows, landslides, scouring or failure of granular bulk carried by vessels.
This paper aims at a two-phase granular flow model supported by the Drucker-Prager yield criterion within a FV-VoF fluid solver. A Bingham fluid approach is combined with an Eulerian strain measure to validate the failure area of granular dam slides. The monolithic framework is chosen to apply the method to real-life problems featuring large displacements. The article carefully examines the performance of the rigid-perfectly plastic material model.

A crucial part of granular simulation models refers to identifying the transition to plastic behavior from a yield criterion. In soil mechanics, a range of yield criteria are used to capture the response of granular material under different conditions. Employed criteria comprise well-known theories, including the Mohr-Coulomb  \cite{Coulomb1776}, Drucker-Prager  \cite{DruckerPrager1952}, Rankine  \cite{Rankine1857}, Tresca  \cite{Tresca1864}, and the von Mises criterion \cite{vonMises1913}. Traditionally, they are incorporated in elasto-plastic models, as, e.g., described in Borja \cite{Borja2013}. The two most frequently used approaches are the Mohr-Coulomb and the Drucker-Prager criterion. The primary challenge associated with the Mohr-Coulomb criterion is the derivation of a general three-dimensional stress state expression for the deviatoric yield stress. Without further simplifications, this often leads to fairly complex formulations, as exemplified in works such as Abbo et al. \cite{AbboSloan1993} and Schajer \cite{Schajer1998}. To circumvent the necessity of calculating principal stresses, formulating an invariant-based yield criterion is highly desirable. Therefore, the Drucker-Prager criterion is preferred in this work. In addition, the Bingham-fluid plasticity approach is used, and we assume an incompressible rigid-perfectly plastic material behavior.

Earlier approaches employing Bingham-fluid plasticity to model granular flow include the work of Mei et al. \cite{Mei1987}, Papanastasiou \cite{Papanastasiou} and Huang et al. \cite{Huang1999}. More recently, similar rheology models were employed in combination with different numerical methods, i.e. Finite Element  (FEM; \cite{zienkiewicz2000finite}), Smooth Particle Hydrodynamics (SPH; \cite{Liu2003}) and  particle Finite Element  (pFEM; \cite{onate2004}) methods, by Leppert \cite{Leppert2007}, Larese \cite{Larese2013}, Ulrich \cite{Ulrich2013} and Ulrich et al. \cite{UlrichEtAl2013}. Wu et al. \cite{Wu2013, Wu2014}, simulated a debris flow  with a Bilinear-Visco-Plastic rheology model inside an industrial FV solver without validating the resulting failure behavior of the sliding material.
Recent literature dealing with two-phase modeling of water-soil interactions includes studies reported by Wang et al. \cite{Wang2016, Wang2021}, who employed a coupled FV-SPH model for simulating a submerged granular column collapse. In these studies, the non-cohesive dam break validation case introduced by Bui \cite{Bui2008} is employed for validating the predicted final shape of the soil surface. The employed SPH model coincides mainly with the elastoplastic model introduced in Bui \cite{Bui2008} and Bui et al. \cite{Bui2011, Bui2017}.  Nguyen et al. \cite{Nguyen2023} utilized a coupled FV-LPT (Lagrangian Particle Tracking) model for simulating the scouring in granular beds. A similar approach to the present paper is explored in Adams et al. \cite{Adams2023}, where turbulence-resolving simulations of dense mud-water mixtures within an FV-VoF employing Bingham plastic materials are verified against simulations by Gavrilov et al. \cite{Gavrilov2017}. Over recent years, the Discrete Element Method (DEM) has received considerable attention in the field of granular flow simulations as outlined in El-Emam et al. \cite{ElEmam2021}.

When considering granular problems like embankment failures, landslides, and the failure of the granular bulk on merchant vessels, not only the development of the shape of the granular material is essential, but also the strain evolution inside the granular material. A strain measure is inherently included in elastoplastic models as, for example, suggested by Bui and co-workers in \cite{Bui2008, Bui2011, Bui2017}, but not included in the FV-VoF approaches mentioned above. An Eulerian strain measure is introduced to the present FV-VoF approach to fill this research gap, providing valuable insights into the evolving deformation and failure mechanism. The Eulerian displacement equation is solved to compute the strain measure for each time step, providing valuable insights into the evolving deformation.The interface-capturing VoF approach introduced by Hirt et al. \cite{Hirt1981} represents the free boundary between the two incompressible immiscible phases. Utilizing the Drucker-Prager yield criterion within a Bingham fluid formulation applied to the soil phase while concurrently addressing the incompressible Navier-Stokes equations within the air phase culminates in a unified, monolithic approach. Additionally, the Eulerian finite strain measure in the presented Bingham-plasticity approach is used to validate the soil shape and the failure area for non-cohesive and cohesive materials, which expands the applicability of the FV-VoF approaches given in Adams et al. \cite{Adams2023}, Wu et al. \cite{Wu2013}, and Wu et al. \cite{Wu2014}.\\ 
Incorporating granular behavior into a fluid solver based on the FV-VoF methodology, a rigid-perfectly plastic two-phase Drucker-Prager model is introduced and validated for non-cohesive and cohesive material with different angles of repose. The model provides a robust framework for simulating the dynamics of granular materials combined with other computational fluid dynamics applications such as scour formation, wave-seabed interaction, or granular cargo modeling on bulk carriers. 

The paper is structured as follows: Sec. \ref{ThBackgr} introduces the mathematical model. Subsequently, the numerical method and the computational algorithm are described in Sec. \ref{NUM}. Section \ref{2DTestcases} is devoted to validation and application studies. Applications are concerned with the failure of geotechnical dam breaks. The results include the soil surface and failure line shapes as well as field values of the deviatoric strain in comparison with the experimental data and numerical data of SPH simulations. Final conclusions and future directions are outlined in Sec. \ref{Concl}. 
Within the publication, the Einstein summation convention is used for lower-case Latin subscripts. Vectors and tensors are defined with reference to Cartesian coordinates. 

\vspace{-6pt}

\section{Mathematical Model}
\label{ThBackgr}
\vspace{-2pt}
The rigid-perfectly plastic material is modeled within a monolithic two-phase VoF framework, i.e. the density $\rho$ and viscosity $\mu$ follow from a soil volume concentration or mixture fraction field $c$, viz.   
\begin{equation}
\rho=(1-c) \rho^A+ c \rho^S\, ,  \qquad 
\mu=(1-c) \mu^A+ c \mu^S\;,
\label{stoffgesetzDens}
\end{equation}
where the invariable bulk properties are denoted by $\rho^A, \mu^A$ for the air phase and $\rho^S, \mu^S$ for the soil phase. 
Applying the immiscibility condition ($D c$/$D t=0$) as well as the incompressibility condition for the bulk densities $\rho^A$ and $\rho^S$,  the differential continuity equation simplifies  to the usual zero-divergence condition for the velocity $v_i$ 
\begin{equation}
\frac{D \rho}{D t}= - \rho \frac{\partial v_i }{\partial x_i}
  \quad \to \quad 
   (\rho^S -\rho^A) \frac{D c }{D t}
    = - \rho \frac{\partial v_i }{\partial x_i}
    \quad \to \quad 
\frac{\partial v_i }{\partial x_i}=0 \, .
\label{conti}
\end{equation}
The transport of the soil mixture fraction $c=V^{S}$/$V$ -- which determines the soil occupied volume $V^{S}$ inside a (control) volume $V$ -- follows from the Eulerian form of the immiscibility condition modified for $\partial v_i /\partial x_i=0$ , i.e.  
\begin{equation}
\frac{\partial c}{\partial t}+\frac{\partial \left( c v_i \right)}{\partial x_i}=0 \, . 
\label{contifresco}
\end{equation}
To model the rigid-plastic material within a fluid dynamic FV framework, the differential momentum equations of a continuum are formulated in Eulerian coordinates 
\begin{equation}
 \frac{\partial  \rho v_i}{\partial t}+ \frac{\partial \rho v_i v_j}{\partial x_j}=\rho g_i+ \frac{\partial \sigma_{ij}}{\partial x_j} \, , 
 \label{Impulse_2}
\end{equation}
where $\sigma_{ij}$ is the Cauchy stress tensor. By splitting the Cauchy stress into a hydrostatic ($-p\delta_{ij}$) and deviatoric ($\tau_{ij}$) part, the deviatoric Kirchhoff stress tensor $\tau_{ij}$ is expressed as
\begin{equation}
      \tau_{ij}=\mu \left( \frac{\partial v_i}{\partial x_j}+ \frac{\partial v_j}{\partial x_i}\right)=\mu\; 2\dot{\epsilon}_{ij}
      = \big[c \mu^S+(1-c) \mu^A \big]\;2 \dot{\epsilon}_{ij} 
      \, ,
      \label{tauepsilno}
\end{equation}
with $\dot{\epsilon}_{ij}$ being the (traceless) strain rate tensor. The constant bulk viscosity $\mu^A$ yields the classical formulation of the Navier-Stokes equations in the air phase. For the granular phase, a variable viscosity will be introduced by assuming perfect plasticity for the material. The derivation of the variable, isotropic viscosity $\mu^S$ follows from the deviatoric stress tensor inside the granular phase 
\begin{equation}
      \tau^S_{ij}=\mu^S 2 \dot{\epsilon}_{ij} \, . 
      \label{viscousstress}
\end{equation}
The Drucker-Prager yield criterion is chosen as a yield model due to the invariant-based formulation, which supports to a simple derivation for the isotropic viscosity. This corresponds to an idealization of the Mohr-Coulomb criterion where the derivative of the plastic potential is unique at each point on the yield surface. The Drucker-Prager yield criterion is given by
\begin{equation}
\Phi=\sqrt{J_2}+\alpha_{\phi} I_1-k_c
\label{DruckPrag}
\end{equation}
 where $I_1$ is the first invariant of the stress tensor, that relates to the pressure by $I_1=-3p$, and $J_2$ is the second invariant ($J_2=1$/$2\;\tau^S_{ij}\;\tau^S_{ij}$). The Drucker-Prager parameters  $\alpha_{\phi}$ and $k_c$ are determined in accordance with Bui et al. \cite{Bui2008} chosen to enable a model comparison, viz. 
\begin{equation}
\alpha_{\phi}=\frac{tan \phi}{\sqrt{9+12 tan^2 \phi}} \, , 
\end{equation}
\begin{equation}
k_c=\frac{3 C}{\sqrt{9+12 tan^2 \phi}} \, . 
\end{equation}
Here $C$ denotes the cohesion and $\phi$ the angle of repose of the granular material. This gives an approximation of the Mohr-Coulomb criterion assuming plane strain, cf. Neto et al. \cite{Neto2008}. Calculating the second invariant of the deviatoric stress inside the granular phase from Eqn. \eqref{viscousstress}, leads to 
\begin{equation}
J_2=2 \left(\mu^S\right)^2 \dot{\epsilon}_{ij} \dot{\epsilon}_{ij}
\end{equation}
and can be inserted in the Drucker-Prager yield function Eqn. \eqref{DruckPrag}. For perfectly plastic materials, the yield function is always zero, and therefore, by rearrangement of Eqn. \eqref{DruckPrag}, the variable viscosity is obtained
\begin{equation}
\mu^S= \frac{3 \alpha_{\phi} p+k_c}{ \sqrt{2 \dot{\epsilon}_{ij} \dot{\epsilon}_{ij}}} \, . 
\label{muDruckPrag}
\end{equation}
For small strains infinite viscosities are obtained from Eqn. \eqref{muDruckPrag} that mimic the rigid behavior of the granular material. Furthermore, for large strain rates ($\dot{\epsilon}_{ij} \to \infty$) the viscosity approaches zero. This can lead to instabilities in the numerical model. To overcome this problem, a regularized Bingham approach is used 
\begin{equation}
\mu^S=\mu^S_{min}+\frac{3 \alpha_{\phi} p+k_c}{ \sqrt{2 \dot{\epsilon}_{ij} \dot{\epsilon}_{ij}}}\left(1-e^{\left(-m { \sqrt{2 \dot{\epsilon}_{ij} \dot{\epsilon}_{ij}}}\right)}\right)
\label{VarMuLarese}
\end{equation}
 The minimum viscosity $\mu^S_{min}$ has to be as small as possible to obtain perfectly-plastic behavior in Eqn. \eqref{VarMuLarese}. A value of $\mu^S_{min}=10^{-3}$\;Pa\;s is found to give good results for perfectly-plastic material. The constant $m$ in the regularized variable viscosity determines the maximum viscosity. It has to approach infinity to obtain a rigid material behavior; nevertheless, it should be bounded for reasons of numerical stability. For vanishing strains, i.e.  $\sqrt{2 \dot{\epsilon}_{ij} \dot{\epsilon}_{ij}} \to 0$, the maximum viscosity refers to $\mu^S_{max}=m\left( 3 \alpha_{\phi} p+k_c\right)$ and its magnitude should be in the order of $10^8$\;Pa\;s for rigid-plastic material behavior. The factor $m$ can also be used to mimic elastic stress-strain behavior inside the yield domain for materials with a larger elastic domain. A maximum viscosity of the order $10^6$\;Pa\;s is proposed for such materials.

Using a cell-centered FV method, the material properties are computed in the cell center and interpolated to the face centers to compute the stresses (fluxes).  
Applying a simple linear interpolation for the material properties leads to a (too) large reduction of the viscosity in cells along the soil surface. This can induce a mesh and material property-dependent creeping behavior of the granular material. Therefore, a nonlinear interpolation method based on an arctangent function noted in Appendix \ref{NonLinMatProp} is used for the material interpolation for non-cohesive granular materials.

The norm of the Euler-Almansi strain tensor is used to compare the 
present results with results from the literature. However, due to the Eulerian formulation, strains cannot be explicitly computed. 
To identify the deviatoric strain inside the rigid-perfectly plastic material, an Eulerian displacement equation is added to the equation system, from which the deformation gradient $F_{ij}$ follows, which in turn allows to compute the Euler-Almansi strain tensor $e_{ij}$
\begin{equation}
e_{ij}=\frac{1}{2}\left(\delta_{ij}-G_{ik}G_{jk}\right)\, . 
\label{euleralmansi}
\end{equation}
Here, $G_{ij}$ is the inverse of the deformation gradient $F_{ij}$, which is related to the Eulerian displacement by 
\begin{equation}
F_{ij}=\delta_{ij}-\frac{\partial u_i}{\partial x_j}\;.
\end{equation}
The Eulerian displacement $u_i$ is calculated from the material derivative 
of the velocity inside the granular material $v^S_i$
\begin{equation}
  \frac{D u_i}{D t}=v^S_i
\, .
\end{equation}
 The latter is  computed  by multiplication of the velocity $v_i$ with the soil mixture fraction $c$, i.e. $v^S_{i}=c v_i$, leading to following convective transport equation 
\begin{equation}
  \frac{\partial u_i}{\partial t}+\frac{\partial c v_j u_i}{\partial x_j}=c v_i\;.
  \label{displvs}
\end{equation}
Model enhancements, for example, by using an elliptic relaxation of the granular velocity as suggested by Richter et al. \cite{Richter2010}, are not included here since the Eulerian displacement $u_i$ is a passive variable that is only used to determine the Euler-Almansi strain $e_{ij}$ and is not fed back into the constitutive model.

\section{Numerical Method}
\label{NUM}
A segregated FV algorithm is applied to all transport equations \cite{Ferziger}. 
 The procedure uses 
 a cell-centered, co-located variable arrangement on unstructured grids to discretize the integral forms of the continuity  \eqref{ContiFacesEqu}, momentum   \eqref{MomEqGranMat}, soil mixture fraction  \eqref{SoilMixtFracEq} and displacement equations \eqref{EQDisplFV} 
\begin{equation}
\oint_{A} v_i \, dA_i=0 \, , 
\label{ContiFacesEqu}
\end{equation} 
\begin{equation}
\int_{V} \frac{\partial}{\partial t} \left(\rho v_i\right) dV+\oint_{A} \left(\rho v_i v_j \right) dA_j=
- \oint_{A} p \,  dA_i+\int_{V} \rho g_i dV+ \oint_{A} \mu \left(\frac{\partial v_i}{\partial x_j}+\frac{\partial  v_j}{\partial x_i}\right) dA_j \, , 
\label{MomEqGranMat}
\end{equation}
\begin{equation}
 \int_{V} \frac{\partial c}{\partial t} dV+ \oint_{A}  c v_i dA_i=0
 \, , \label{SoilMixtFracEq}
\end{equation}
\begin{equation}
 \int_{V}  \frac{\partial u_i}{\partial t} dV + \oint_{A} c v_j u_i \; dA_j= \int_{V}  c v_{i} dV\;.
 \label{EQDisplFV}
\end{equation}
Integrals are approximated through a second-order accurate mid-point rule. An implicit first-order accurate temporal discretization method is employed.  The approximation of convective fluxes in
\eqref{MomEqGranMat}
 and \eqref{EQDisplFV} 
follows from a flux blending scheme, in which 70\% of the method leverages the precision of second-order central differencing. The approximation of the convective term in \eqref{SoilMixtFracEq} utilizes a QUICK scheme, initially introduced by Leonard \cite{Leonard1979}. Diffusive fluxes are obtained from central differences. 
The algorithm employs a SIMPLE-type pressure correction scheme. The pressure correction equation is detailed by Eqn. \eqref{presscorr} in App. \ref{DiscrEq} and largely follows Ferziger \cite{Ferziger}, with details given in Yakubov et al. \cite{Yakubov2015} and V\"olkner et al. \cite{Svenja2017}. The resulting discrete forms of the equation system are given in Appendix \ref{DiscrEq}.
 
Alg. \ref{ProcedureFreSCo} outlines the unsteady two-phase procedure. Within each time step, the sequence of governing equations is iterated to convergence until a predefined residual threshold is obtained. 

\begin{algorithm}
\caption{Incompressible two-phase flow with rigid-perfectly plastic material}\label{euclid}
\begin{algorithmic}[lines]
\State{Initialize $v^0_i$, $p^0$, $c^0$, $u^0_i$, $\rho^0$, $\mu^0$, $e^0_{ij}$}
\State{$n=0$}
\State{$m=0$}
\While{$n \le$ max. number time steps}\Comment{$n$ denotes the current time step}
\State $n=n+1$
\While{residual $<$ residual-threshold}\Comment{$m$ denotes the number of outer iterations}
\State $m= m+1$
\State update properties \Comment{Eqn. \eqref{VarMuLarese}, Eqn. \eqref{stoffgesetzDens}}
\State solve momentum equations \Comment{Eqn. \eqref{MomEqGranMat}}
\State solve first stage of pressure correction equation \Comment{Eqn. \eqref{presseqalgsys}}
\State correct pressures, fluxes and velocities \Comment{Eqns. \eqref{corrVel}, \eqref{corrFlux}}
\State solve second stage of pressure correction equation \Comment{Eqn. \eqref{press2stage}}
\State correct pressures \Comment{cf. Appendix \ref{DiscrPress}}
\State solve soil mixture fraction equation \Comment{Eqn. \eqref{SoilMixtFracEq}}
\State solve displacement equations  \Comment{Eqn. \eqref{EQDisplFV}}
\State calculate Euler-Almansi strain \Comment{Eqn. \eqref{euleralmansi}}
\EndWhile\label{euclidendwhile}
\State{$v^n_i$, $p^n$, $c^n$, $u^n_i$, $\rho^n$, $\mu^n$, $e^n_{ij}$ $\to$\; $v^{n-1}_i$, $p^{n-1}$, $c^{n-1}$, $u^{n-1}_i$, $\rho^{n-1}$, $\mu^{n-1}$, $e^{n-1}_{ij}$}
\EndWhile\label{secondwhile}
\State update properties \Comment{Eqn. \eqref{VarMuLarese}, Eqn. \eqref{stoffgesetzDens}}
\State \textbf{return} $v^n_i$, $p^n$, $c^n$, $u^n_i$, $\rho^n$, $\mu^n$, $e^n_{ij}$ \Comment{defined at output time steps}
\end{algorithmic}
\label{ProcedureFreSCo}
\end{algorithm}

\section{Validation and Application Studies}
\label{2DTestcases}
To assess the behavior of the rigid-plastic material model and its VoF-based FV implementation, two-dimensional dam break cases are studied in comparison with experiments and simulations \cite{Bui2008}. The first study uses experimental data for a non-cohesive artificial material to validate the final static configuration of the granular material model. The granular model is subsequently more rigorously compared with an elastoplastic Smoothed Particle Hydrodynamics (SPH) model proposed by Bui et al. \cite{Bui2008} for a wider range of material properties, including cohesive material. The aim of the analysis is to shed light on the transient behavior of the implemented granulate model and to confirm its accuracy in representing the material behavior under different angles of repose.
Reported results are restricted to the soil surface, numerically extracted at $c=0.5$ in the current model, as well as the failure line, that numerically refers to $|| e_{ij} ||=0.3$ for the present model. 
Note that the  case investigated in Sec. \ref{BuiVarPhi} refers to realistic dimensions and material properties, thereby increasing the transferability of the results into practical applications. 

\subsection{Non-cohesive 2D Dam Breaks}
\label{SoilCollapse}
\begin{figure}[tb]
\centering
 \includegraphics{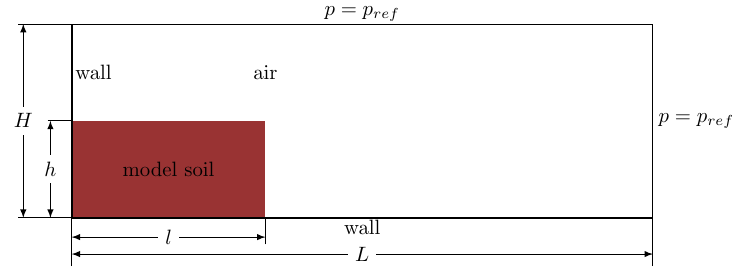}
\caption{Initial configuration and boundary conditions of the 2D soil collapse case (all dimensions in meter).}
\label{2DBuiInit_x_}
\end{figure}

Bui et al. \cite{Bui2008} performed a series of experiments on the collapse of a two-dimensional soil to validate numerical simulations. The two-dimensional environment is emulated by employing aluminum bars with diameters of $1$\;mm and $1.5$\;mm, each possessing a length of $50$\;mm, to simulate the soil material. As outlined in Fig. \ref{2DBuiInit_x_}, the initial configuration is a rectangular enclosure measuring $l=0.2$\;m in width and $h=0.1$\;m in height, with both sides enclosed by retaining walls to secure the aluminum bars in position. The collapse is initiated by the swift removal of a gate at the right end. The material density of the simulated soil is recorded at $2650$kg/m$^3$. Parameters for the constitutive equation are obtained through shear box tests, revealing that the aluminum bars exhibit a complete absence of cohesion. Furthermore, the friction angle $\phi$ is empirically determined to be $19.8^{\circ}$. The Poisson's ratio was estimated to a value of approximately $0.3$ and allows the quantification of the elastic bulk modulus $K$, which is found to be around $0.7$\;MPa.\\
The employed rigid-perfectly plastic granular material model requires only two closure parameters to be determined, i.e. the cohesion $C$ and the internal friction angle $\phi$ which are set to $C=0$\;kPa and $\phi=19.8^\circ$, respectively. The arctangent interpolation  \eqref{NewIntCorr} of both, the viscosity and density is applied in combination with $M=-0.45$ and $N=25$ to suppress soil creeping along the left wall. The regularization parameter of the Bingham model (\ref{VarMuLarese}) is set to $m=10^{5}$ which returns $\mu^S_{max}= 8.6 \cdot 10^7$ \;Pa\;s.

\subsubsection{Validation against Experimental Data}
\label{sec:bui-val}
The computational domain spans $0.6$\;m in length $L$  and $H=0.2$\;m in height. The left and bottom boundaries of the domain are configured as a wall. The top and right far field boundaries are assigned to atmospheric pressure boundaries. No mobile gate is included in the computation and the simulation begins with the collapse. A Cartesian spatial mesh featuring a grid spacing of $\Delta x_1=\Delta x_2=L$\;/$300$ is employed. The uniform time step is assigned to $\Delta t=10^{-4}= 20\; \Delta x_1/\sqrt{g h}$\;s. The initial pressure within the artificial soil refers to the hydrostatic pressure. 

\begin{figure}[tb]
\centering
 \includegraphics{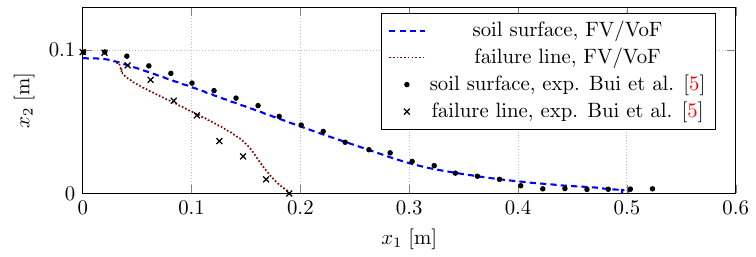}
\caption{Comparison of the measured and predicted final surface and failure line for the validation case of Bui et al. \cite{Bui2008} ($t=20$\;s; $\phi = 19.8^\circ$;  $C=0$).}
\label{BuiOhneInterp_MeshStudyconstN}
\end{figure}

 Fig. \ref{BuiOhneInterp_MeshStudyconstN} 
compares the results obtained from the present simulations with experimental data of  Bui et al. \cite{Bui2008} at a final point in simulation time. The figure depicts the soil surface, numerically extracted at ($c=0.5$), as well as the failure line ($|| e_{ij} ||=0.3$). It has to be noted that the displayed failure line of the simulation is cut at the intersection with the soil surface.  
A close agreement of the soil surface with experimental data is observed. The greater drop in column height at the left end is due to the reduced friction forces associated with the chosen simple interpolation on the free surface.
This also results in slightly lower angles of repose near the vertical wall. The failure line of the rigid-plastic model agrees less well with the experimental data, which is attributed to the simplifying rigid-perfectly plastic material model. 
\begin{figure}[tbp]
  \centering
  \includegraphics[scale=.48]{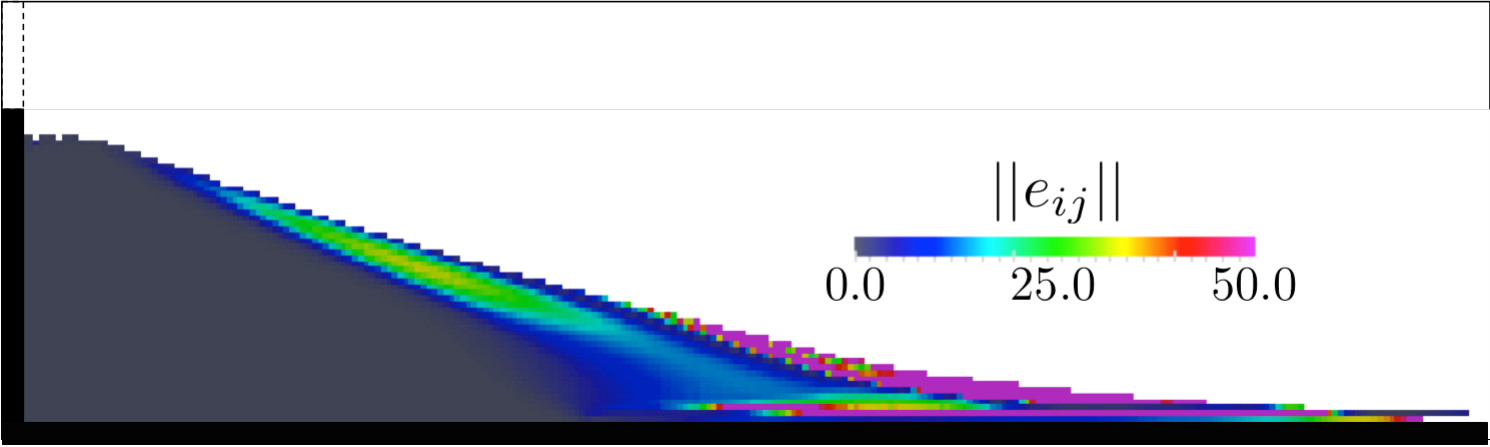}
  \includegraphics[scale=.8]{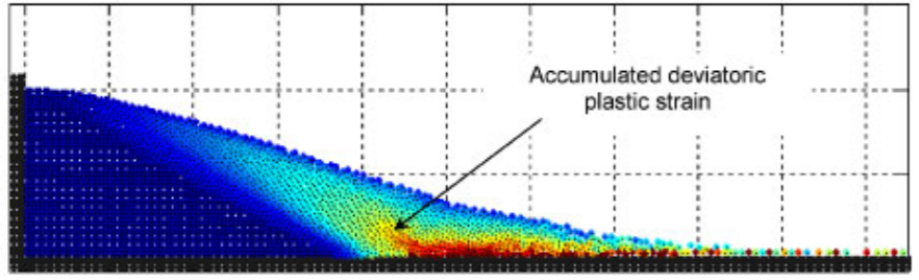}
  \caption{Comparison of strain predictions for a final time  ($t=20$\;s) of the 2D  validation case of Bui et al. \cite{Bui2008} ($\phi = 19.8^\circ$;  $C=0$). Top: Norm of the deviatoric Euler-Almansi strain predicted by the current method. Bottom: Accumulated deviatoric strain from \cite{Bui2008}.}
  \label{BetrEulerAlmansiBui}
\end{figure}

In order to discuss the quality of the material model in more detail, Fig. \ref{BetrEulerAlmansiBui} compares the norm of Euler-Almansi strain tensor $||e_{ij}||$ returned by the present rigid-plastic model (top) with the accumulated deviatoric strain field of the reference elastoplastic model utilized in Bui et al. \cite{Bui2008} (bottom). Though no quantitative comparison is possible, since no legend is provided in Bui et al. \cite{Bui2008}, an enhanced strain level is observed at the lower right boundary in both simulations. The considerable accumulation of strain at the soil surface depicted in the present simulation, follows from the Eulerian displacement computation \eqref{displvs}. The phenomenon arises due to the multiplication of the velocity field $v_i$ by the soil mixture fraction, leading to a significant displacement gradient and related deformation gradient $F_{ij}$ at the soil-air interface. The elevated $F_{ij}$ values are aligned with elevated Euler-Almansi strains.
The model could be improved by a nonlinear face-interpolation relation for the soil velocity or introducing an elliptic relaxation equation for the soil velocity prediction. However, this is omitted here, since the deformation gradient is a post-processing aspect and does not affect the material behavior. Applying a Cahn-Hilliard-based two-phase flow approach as suggested in K\"uhl et al. \cite{Kuehl2021a} could lead to independence of the interpolation functions for interphase properties with respect to the chosen constants.

\subsubsection{Verification of Transient Behavior}
\label{BuiVarPhi}
The second case is concerned with the verification of the transient behavior 
of the present computational model in comparison to SPH results reported in Bui et al. \cite{Bui2008} for an elastoplastic material model. Displayed results correspond to three distinct angles of repose, i.e. $\phi = 25^\circ$, $45^\circ$ and $65^\circ$.   
The configuration corresponds to the previously examined scenario depicted in Figure \ref{2DBuiInit_x_}. However, the dimensions have been expanded, i.e. the initial dam length and height read  $l=4$\;m and  $h=2$\;m, and the computational domain spans length $L=12$\;m and a height of $H=4$\;m. To ensure a  realistic context, a soil density of 1850 kg/m $^3$ is used and the general dimensions are similar to practical applications.

A homogeneous Cartesian spatial mesh is employed, which is again characterized by $\Delta x_1=\Delta x_2=L$\;/$300$. Moreover, a uniform time step of $\Delta t=10^{-4} = 89\; \Delta x_1/\sqrt{g h}$\;s is applied. The initial and boundary conditions remain consistent with those outlined in Section \ref{sec:bui-val}. In order to prevent deviations in the first time step, the material density is initially set to $20.0$\;kg/m$^3$ at the beginning of the simulation. The tenth time step then switches to $1850$\;kg/m$^3$. This dynamic density adjustment is required by the increased static pressure within the soil column. 

\begin{figure}[h]
  \centering
   \subfigure{ 
\includegraphics{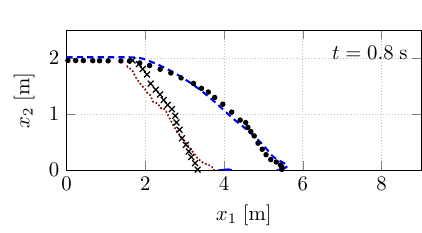}
}
   \subfigure{ 
\includegraphics{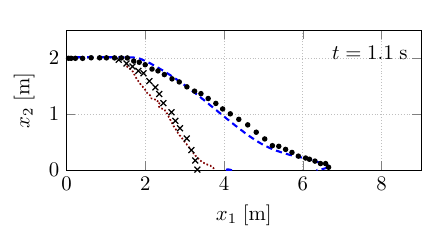}
   }
      \subfigure{ 
\includegraphics{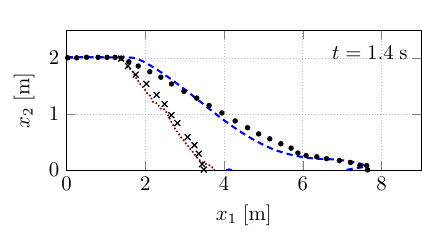}
   }
      \subfigure{ 
\includegraphics{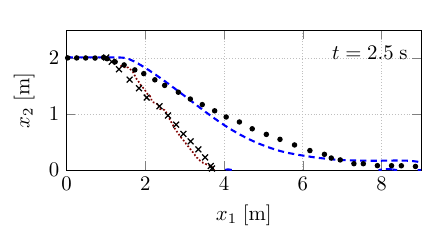}
   }
\includegraphics{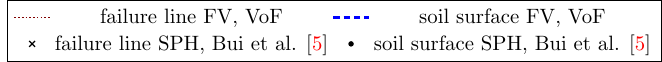}
     \caption{Comparison of the temporal evolution of the soil surfaces and failure lines predicted for the 2D soil collapse verification case at $\phi = 25^\circ$ angle of repose  and $C=0$ with two different computational models.  
     }
  \label{phi25Bui}
\end{figure}

\begin{figure}[h]
  \centering
    \subfigure{ 
\includegraphics{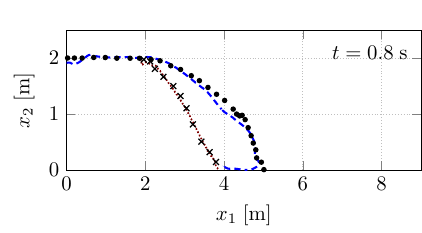}
}
      \subfigure{  
\includegraphics{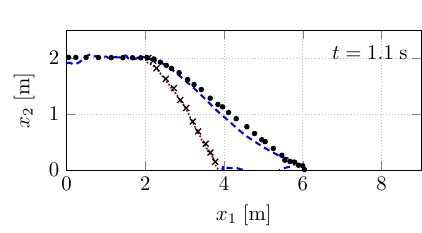}
}
      \subfigure{ 
\includegraphics{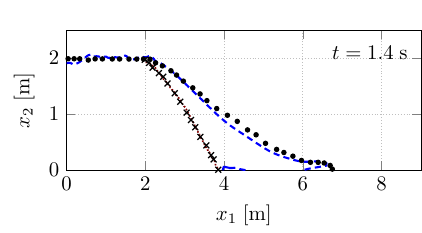}
}
     \subfigure{ 
\includegraphics{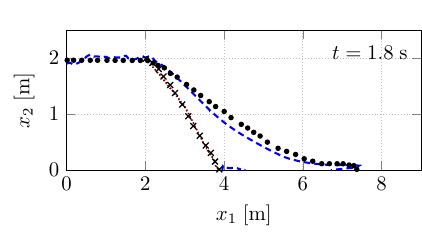}
}
\includegraphics{BuiPhi25_legend.pdf}
    \caption{
    Comparison of the temporal evolution of the soil surfaces and failure lines predicted for the 2D soil collapse verification case at $\phi = 45^\circ$ angle of repose and $C=0$ with two different computational models.}
  \label{phi45Bui}
\end{figure}

\begin{figure}[h]
  \centering
    \subfigure{ 
\includegraphics{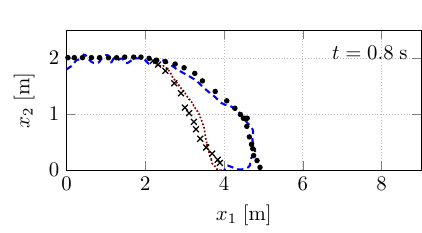}
}
    \subfigure{ 
\includegraphics{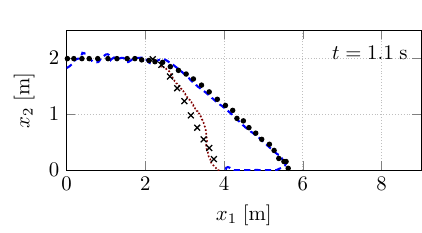}
}
   \subfigure{ 
   \includegraphics{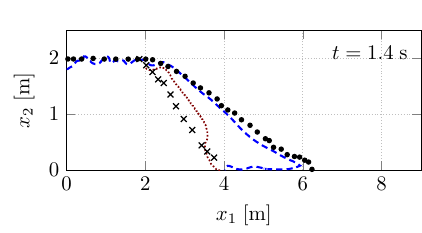}
}
   \subfigure{ 
\includegraphics{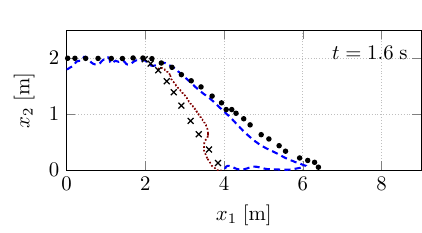}
}
\includegraphics{BuiPhi25_legend.pdf}
    \caption{Comparison of the temporal evolution of the soil surfaces and failure lines predicted for the 2D soil collapse verification case at $\phi = 65^\circ$ angle of repose and $C=0$ with two different computational models.}
  \label{phi65Bui}
\end{figure}

Figs. \ref{phi25Bui}, \ref{phi45Bui}, and \ref{phi65Bui} portrait the simulation results for the three investigated repose angles in combination with zero cohesion $C=0$.  Results of the computed soil surface and failure line are compared between the present approach and SPH simulations results reported by Bui et al. \cite{Bui2008}, which employed
a particle diameter close to the present grid spacing $\Delta x_1$ and are indicated with symbols. The comparisons refer to four time instants $t=0.8$\;s, $t=1.1$\;s, $t=1.4$\;s and an additional, variable fourth time instant. 
Again, only the diagonal part of the failure line returned by the current model is displayed, which is related to an Euler-Almansi strain norm of $0.3$.
Since the hydrostatic pressure is higher compared to the previous case,  higher yield stresses occur. Therefore, although the angle of repose is similar, the results are expected to differ. The diagonal part of the failure line in Fig. \ref{phi25Bui} agrees well with the SPH results of Bui et al. \cite{Bui2008}, with minor compromises for the first time step. 
The soil surface is well represented at $t=0.8$\;s and $t=1.1$\;s. However, it then becomes more convex than the reference SPH solution. At $t=2.5$ s, the surface is higher for $x_1>8.0$\;m due to the numerical diffusion inherent to the approximation of soil mixture fraction convection at the soil surface. 

When attention is directed to the $45^\circ$ angle of repose, it becomes evident that distinct settings for the interpolation function and the regularization parameter $m$ are required. This requirement arises from the notably increased yield stress exhibited by materials with a steeper angle of repose, which motivates the reduction from $m=10^{5}$ to $m=10^{4}$ to obtain a sufficiently large maximum viscosity of $\mu^S_{max}\ge 10^8$\;Pa\;s and mimic the elastic behavior for small strains. Due to the increase in the size of the yield domain, the elastic material behavior in the SPH results plays a larger role than in the previous case.
A linear interpolation of viscosity and density is applied to ensure a smooth transition between the properties of the soil and the surrounding air. The results for this particular angle of repose are depicted in Fig. \ref{phi45Bui}. A good agreement of the failure line is observed, particularly in its diagonal section where the reference SPH solution and the present model agree very well.
 The current model accurately represents the steep gradient of the soil surface at $t=0.8$\;s. Again, the agreement of the predicted soil surfaces deteriorates at later times and the current model manifests a more convex shape of the compared to the reference SPH  results. These differences are likely attributable to the substantially different constitutive soil models.
  Furthermore, a minor disturbance in the soil surface along the left wall can be observed, which relates to creeping phenomena and does not increase over time. Another concern arises at the bottom wall where air inclusion is observed at $x_1=4.0$\;m. This phenomenon can be attributed to the exceedingly high viscosity in the lower right corner of the original dam geometry, which results from elevated pressure values. 

Moreover, an increase in the angle of repose initially leads to a greater velocity of the lower surge front in the SPH simulations, see Fig. \ref{phi45Bui}. This difference becomes even more evident at the largest slope angle examined here, cf. Fig. \ref{phi65Bui}.
 In contrast, the large hydrostatic pressures in the current model results in large viscosities and associated wall shear forces, which are obviously higher than the wall shear force predicted by the SPH model. The reasons for this may lie in the material modeling, but can also be very much related to the fundamentally different implementation of the wall influences in SPH, where a where a uniform stress state is enforced.  
 The significantly smaller soil velocities predicted by the current model induce small air inclusions, and it is also expected that incorporating an elastoplastic model will address and resolve this issue by reducing the wall-shear forces. 

In the case of an angle of repose of $\phi = 65^\circ$, the regularization parameter $m$ is reduced to $m=10^{2}$ to emulate elastic behavior inside the increased yield domain. The maximum viscosity is obtained as $\mu^S_{max}= 2.9 \cdot 10^6$\;Pa\;s. At the same time, the linear interpolation technique for density and viscosity is maintained. As illustrated in Fig. \ref{phi65Bui}, the challenges observed for an angle of repose of $\phi = 45^\circ$ are exacerbated when dealing with  $\phi = 65^\circ$. This can be attributed to the substantially elevated yield stress associated with the steeper angle of repose, which, in turn, necessitates the representation of larger viscosity values. In the rigid section of the dam ($x_1<2$\;m), the soil surface predicted by the current approach exhibits a notably undulated profile, a phenomenon arising from the augmented stress gradient at the soil surface in granular materials featuring increased yield stress. Although the alignment of the failure line with the reference data is far less precise than observed for materials with $\phi = 45^\circ$, it is still deemed satisfactory within the scope of the present model.
It is important to note that as the plastic yield stress increases, the challenge of accurately simulating granular materials increases for the rigid, perfectly-plastic FV-VoF method, indicating a need for an elastoplastic formulation.

\subsection{Cohesive 2D Dam Break}
In order to assess the performance of the suggested computational soil model under cohesive material conditions, the $\phi = 25^\circ$ case presented in Sec. \ref{BuiVarPhi} is investigated for a cohesive material with $C=5$\;kPa. Boundary conditions, dimensions, and numerical grids for this verification case agree with the corresponding case in
Sec. \ref{BuiVarPhi}. Introducing cohesion effectively increases the yield stress, similar to the $\phi = 65^\circ, C=0$ case, and the regularization parameter is again set to $m=10^2$ to emulate elastic behavior inside the yield domain ($\mu^S_{max} = 1.9 \cdot 10^6$). The interpolation constants $M$ and $N$ are defined as $M=0$ and $N=5$ for this investigation. 

\begin{figure}[htbp]
  \centering
    \subfigure{ 
\includegraphics{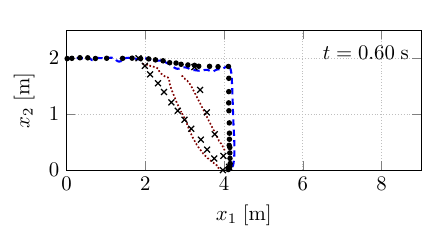}
}
  \subfigure{ 
\includegraphics{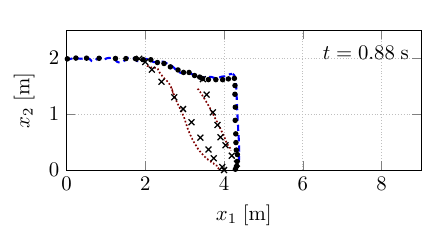}
}
  \subfigure{ 
\includegraphics{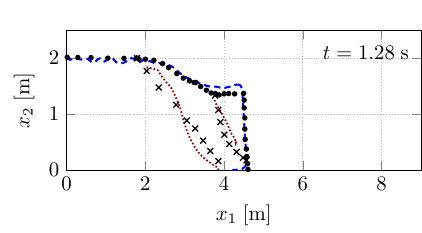}
}
  \subfigure{ 
\includegraphics{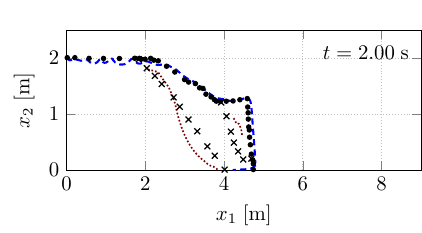}
}
\includegraphics{BuiPhi25_legend.pdf}
    \caption{Comparison of the temporal evolution of the soil surfaces and failure lines predicted for the 2D soil collapse verification case at $\phi = 25^\circ$ angle of repose and $C=5$ with two different computational models.}
 \label{BuiCoh}
\end{figure}

 Results of the current simulations are compared to corresponding results reported by Bui et al. \cite{Bui2008}. 
 Fig. \ref{BuiCoh}, again displays the predicted soil surfaces and failure lines for four different time instants. Notably, the cohesion-induced augmented yield stress introduces indentations of the soil surface along the rigid portion of the dam in both simulations. Additionally, a thin layer of air is observed between the granular material and the bottom wall at later time steps for the current simulation. However, apart from these observed nuances, the soil surface of the current model is largely in line with that of the SPH reference model. This confirms the general functionality of the presented model when considering cohesive materials.
For a better comparison with the SPH results, the failure lines are only partially displayed, and failure lines close and parallel to the surface are omitted. Consequently, the rigid portion of the granular material is aptly represented, even at later time intervals. The failure lines depicted in Fig. \ref{BuiCoh} exhibit a remarkable agreement with the data derived from the SPH model. Minor variations are expected, primarily attributable to disparities in the underlying material models. Nevertheless, the results of the comparison demonstrate the good performance of the presented model when dealing with cohesive materials.
\begin{figure}[htbp]
\centering
  \subfigure[$t=0.6$\;$\si{\second}$]{ 
    \includegraphics[scale=0.185]{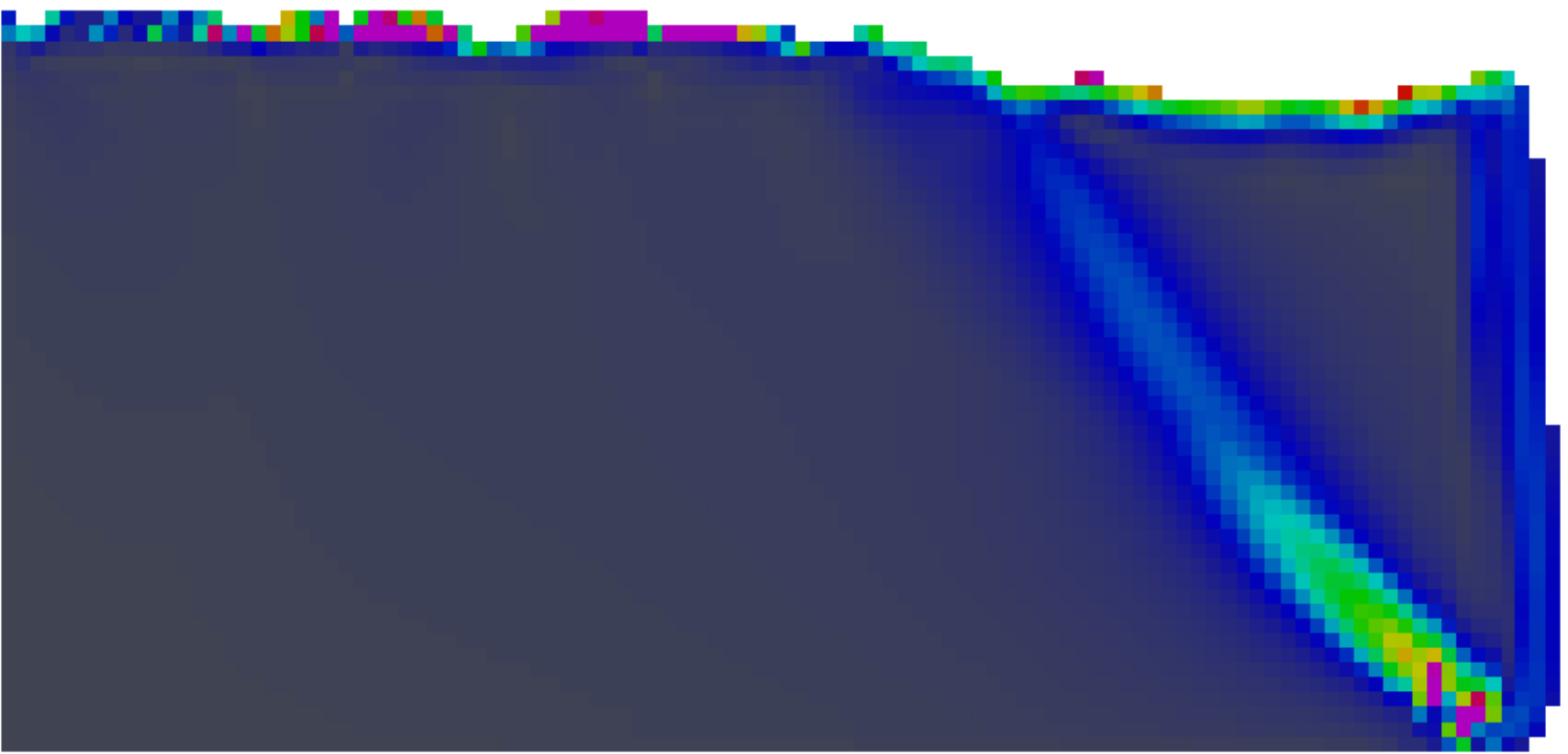} \hfill
}
 \subfigure[$t=0.6$\;$\si{\second}$]{ 
   \includegraphics[scale=0.6]{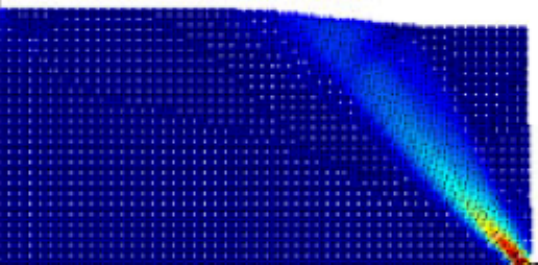}
}
  \subfigure[$t=0.88$\;$\si{\second}$]{ 
   \includegraphics[scale=0.19]{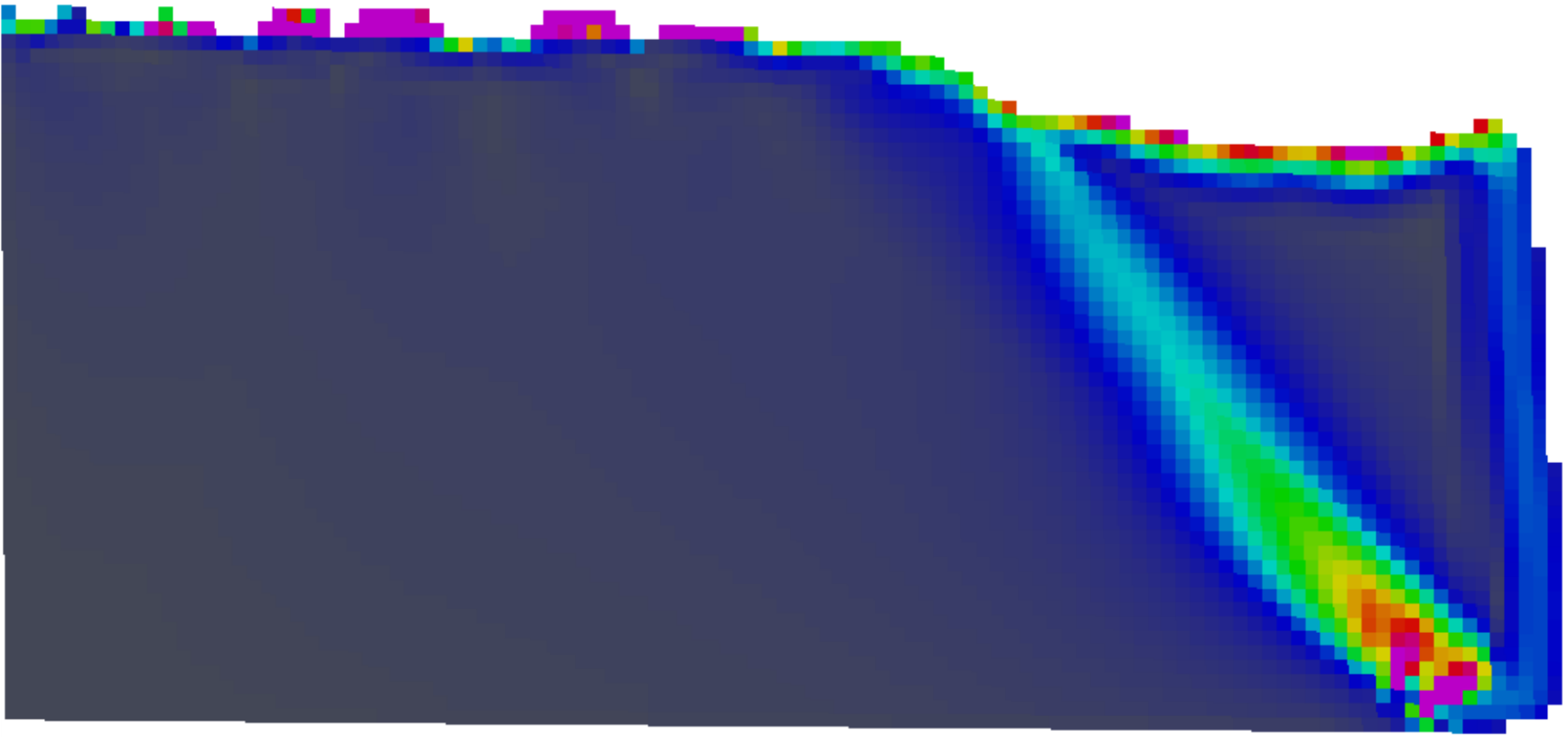} \hfill
}
 \subfigure[$t=0.88$\;$\si{\second}$]{ 
  \includegraphics[scale=0.6]{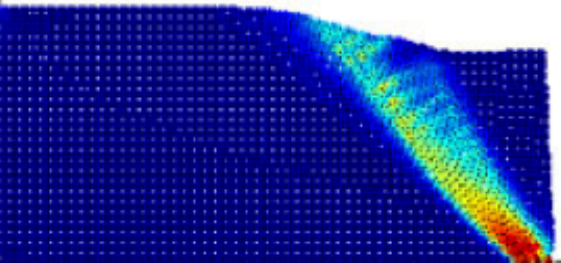}
}
 \subfigure[$t=1.28$\;$\si{\second}$]{ 
   \includegraphics[scale=0.195]{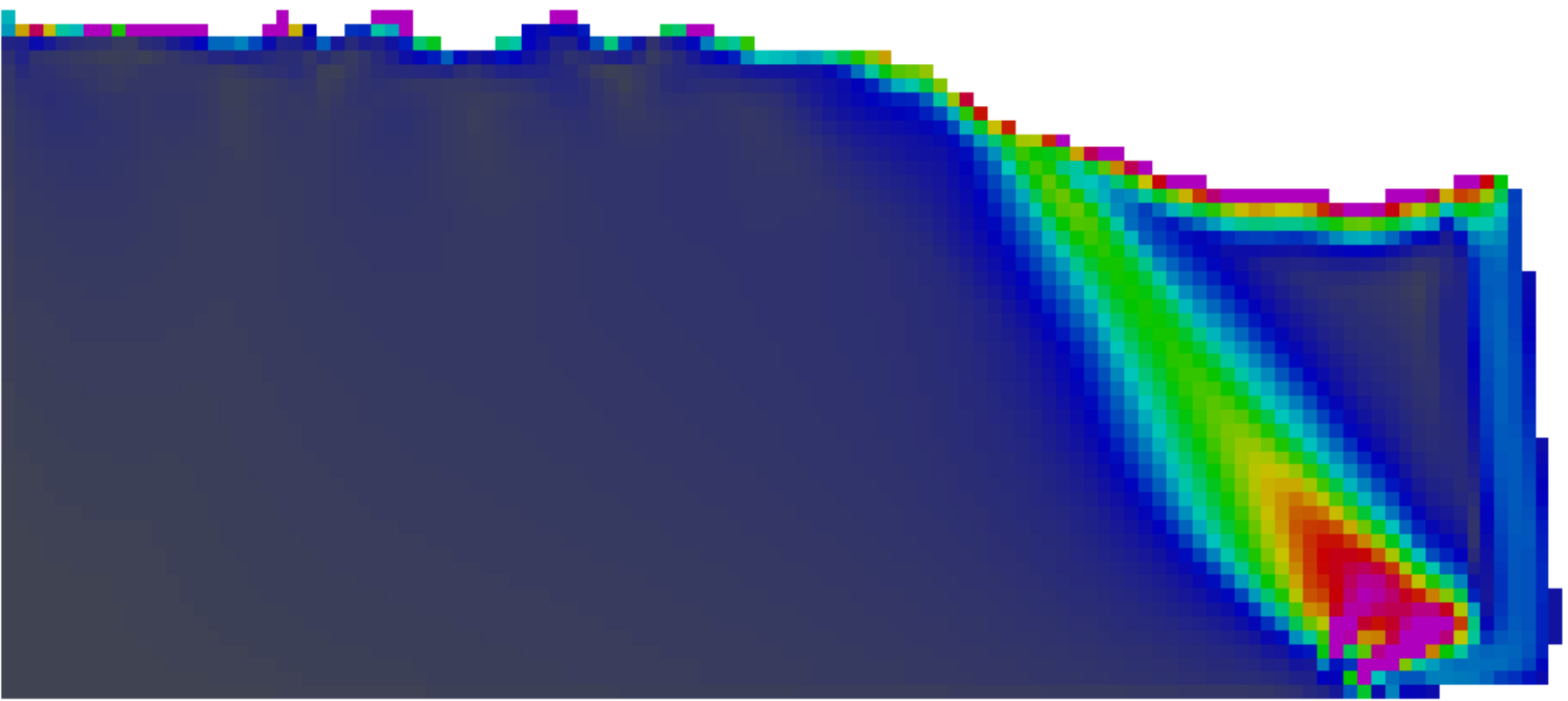}
}
 \subfigure[$t=1.28$\;$\si{\second}$]{ 
   \includegraphics[scale=0.6]{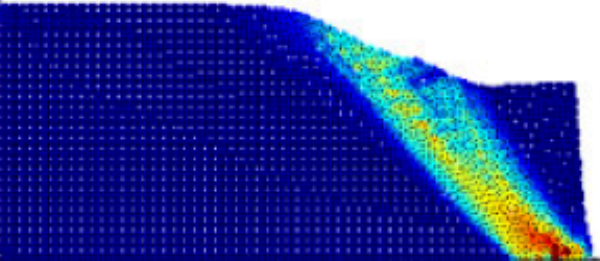}
}
 \subfigure[$t=2.00$\;$\si{\second}$]{ 
        \includegraphics[scale=0.21]{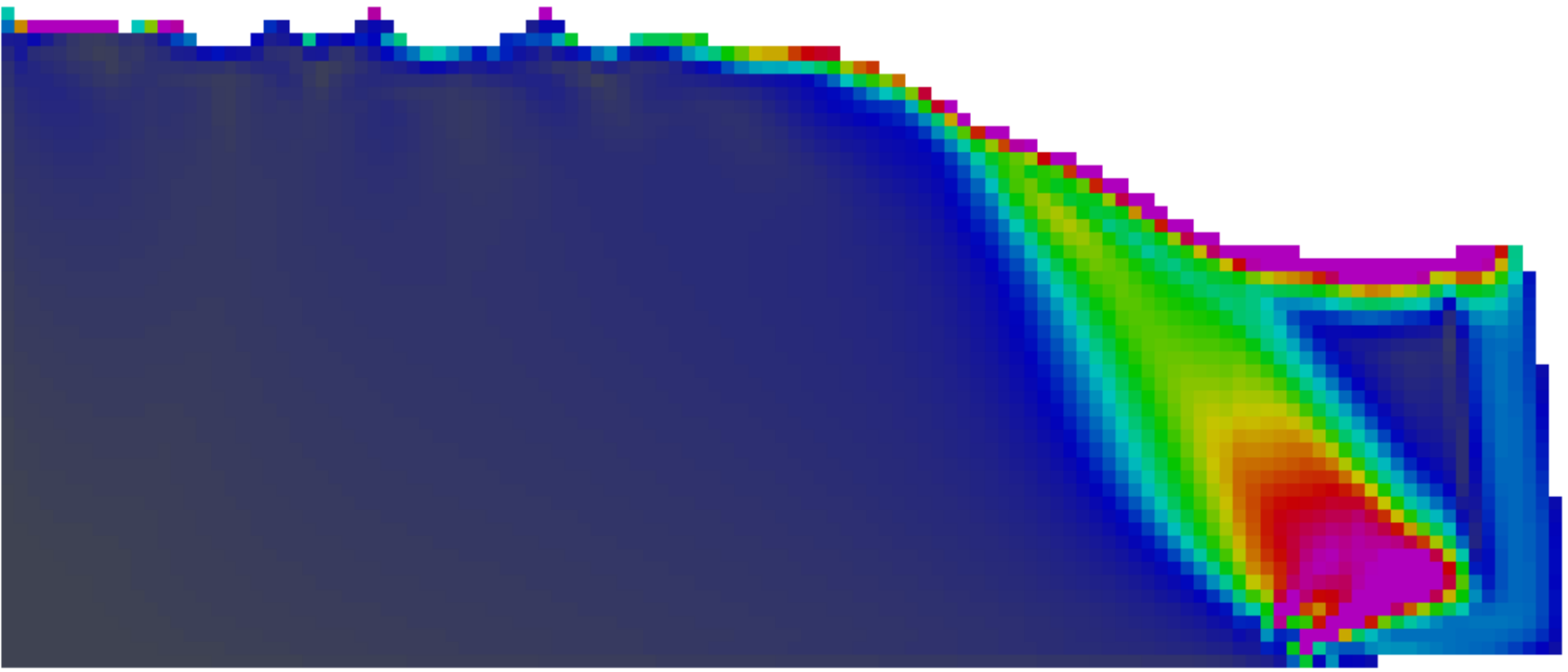}
}
 \subfigure[$t=2.00$\;$\si{\second}$]{ 
   \includegraphics[scale=0.6]{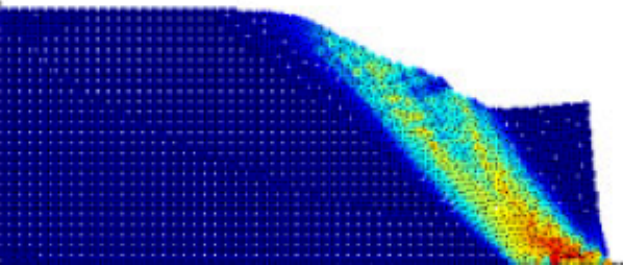}
}
  \includegraphics[scale=0.6]{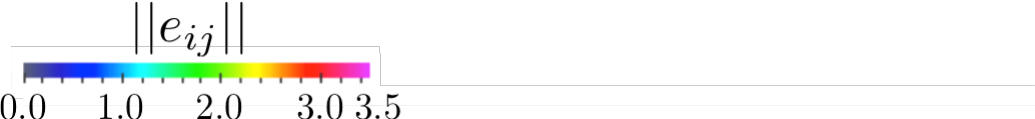}
    \caption{Comparison of  predicted strain evolutions of the 2D cohesive soil collapse verification case ($\phi = 25^\circ$, $C=5$\;kPa). Left: Norm of the deviatoric Euler-Almansi strain predictied by the current method. Right: Accumulated deviatoric strain from Bui et. al \cite{Bui2008}.}
 \label{BuiCohField}
\end{figure} Fig. \ref{BuiCohField} provides a qualitative comparison of the predicted norms of the Euler-Almansi strain tensor (left) and absolute values of accumulated deviatoric strain (right), respectively. 
 Except for the areas near the soil surface, the comparison shows a high degree of agreement. For the current rigid perfectly-plastic FV model, the error cone for the initial snap shots ($t=0.6$\;s, $t=0.88$\;s, and $t=1.28$\;s) has a larger width than that of the elastoplastic SPH model, especially at the top of the dam.
 Additionally, the SPH model predicts a less accumulated strain in the elastic right corner of the dam, in contrast to the Euler-Almansi strain norm observed in the rigid portion of the FV model. Deeper inside the granular phase, where $c > 0.99$, the Euler-Almansi strain norm closely aligns with the accumulated deviatoric strain results derived from the SPH model by Bui et al. \cite{Bui2008}.
 
\section{Conclusion}
\label{Concl}
A rigid-perfectly plastic Bingham model is presented and its implementation into a FV-VoF procedure is validated for cohesive and non-cohesive materials featuring different angles of repose. A close agreement of the predicted soil surface with experimental data is obtained for non-cohesive material. 
To verify the applicability to realistic problems, the current procedure was successfully verified in large-scale dimensions against SPH simulations that use a more sophisticated material model.
As the yield stress increases with increasing angle of repose and cohesion, the challenge of accurately simulating granular materials becomes ever greater for the current  method.
Nevertheless, the soil surface shape and strain results inside the granular material agree surprisingly well with the SPH results for cohesive verification cases.
The advantage of the method is the ability to simulate large displacements and arbitrary soil shapes with a standard FV-VoF solver. 
The failure area can be computed from the introduced Euler-Almansi strain measure, which extends the scope to geotechnical problems such as landslides.

Challenges arising at the boundary of two-phase flow are discussed and circumvented by introducing a nonlinear interpolation function for the phase field. 
An elliptic relaxation of the soil velocity night further improve the approach and avoid spurious increased values of the deformation gradient tensor.

Future research will consider a monolithic elasto-plastic VoF approach  to extend the applicability of the presented model. Moreover, we will also  include a coupling of the presented method with a free surface flow through porous media, as described in D\"usterh\"oft-Wriggers et al. \cite{DuesWri2022}, to simulate seepage flow through permeable, perfectly-plastic dams.

\section*{Acknowledgements}
The authors acknowledge the support within the research project ``LiquefAction - Cargo Liquefaction in Ship Design and Operation'' (Grant No. FKZ 03SX363A ). 
Selected computations were performed with resources provided by the North-German Super-computing Alliance (HLRN; grant hhi00016). 

\bibliography{BibPerfPlastic}
\bibliographystyle{plain} % We choose the "plain" reference style

\appendix
\section{Non-linear interpolation of material properties}
\label{NonLinMatProp}
The non-linear interpolation based on an arctan function is applied for the material properties density and viscosity. Here, $\gamma$ represents a general property, and $^A$ and $^S$ denote the air and soil phases, respectively. With the constants $N$ and $M$, the shape of the interpolation can be controlled. The employed interpolation function gets
\begin{equation}
\begin{split}
\gamma=\gamma^A+\left(\frac{arctan\left(N\left(c-M-0.5\right)\right)}{\pi}+0.5\right)(\gamma^S-\gamma^A)\\
-\left(1-\left(c-M\right)\right)\epsilon_0\left(\gamma^S-\gamma^A\right)+\left(c-M\right)\epsilon_1\left(\gamma^S-\gamma^A\right)\\
-\left(1-c\right)\delta_0\left(\gamma^S-\gamma^A\right)+T_{corr.}
\end{split}
\label{NewIntCorr}
\end{equation}
with 
\begin{equation}
T_{corr.}=
\begin{cases}
c\;\delta_1\left(\gamma^S-\gamma^A\right), & \text{if}\ M>0 \\
-c\;\delta_1\left(\gamma^S-\gamma^A\right), & \text{if}\ M<0\;.\\
\end{cases}
\end{equation}
and
\begin{equation}
\epsilon_0=\frac{arctan\left(-0.5N\right)}{\pi}+0.5
\end{equation}
\begin{equation}
\epsilon_1=0.5-\frac{arctan\left(0.5N\right)}{\pi}\;.
\end{equation}
as well as 
\begin{equation}
\delta_0=\left(\frac{arctan\left(N\left(-M-0.5\right)\right)}{\pi}+0.5\right)-\left(1+M\right)\epsilon_0-M\epsilon_1
\end{equation}
\begin{equation}
\delta_1=\biggl| \left(\frac{arctan\left(N\left(1-M-0.5\right)\right)}{\pi}\right)-M\epsilon_0-M\epsilon_1-\left(\frac{arctan\left(N\left(1-0.5\right)\right)}{\pi}\right)\biggr|\;.
\end{equation}

\section{Discrete equations}
\label{DiscrEq}
In the present work, a pressure-based FV formulation using a cell-centered, co-located variable arrangement on unstructured polyhedral grids is applied to approximate the two-phase granular flow problem as presented in Alg. \ref{ProcedureFreSCo}. The applied FV method discretizes the equations formulated in an integral form by dividing the computational domain into finite control volumes of size $\Delta V_P$. The cell center of each control volume is denoted $_P$, and the values at the faces $_f$ are needed to discretize the surface integrals. In the following sections, $A$ denotes the face area, and $A_{i}$ is the outward pointing face vector. The scalar distance between the cell center and adjacent neighboring centers $_{NB}$ is labeled $d$, and its vector is $d_i$. A time step size of $\Delta t$ is used, and $g_i$ is the acting acceleration due to gravity.

\subsection{Discretized Momentum Equations}
Using a mid-point integration rule together with the first-order implicit time discretization and an implicit upwind-difference scheme part for the convective flux, the discrete form of the momentum equations \eqref{MomEqGranMat}gets
\begin{equation}
\begin{split}
v^{n,m}_{i, P} \; \left[ \Delta V_P \, \frac{\rho_P}{\Delta t} + \sum_{f(\Delta V_P)} A^{v_i}_{NB}
\right] - \sum_{f(\Delta V_P)}
\underbrace{\left[ \left(\; max \left \lbrack - \dot{m}^{n,m-1},0 \right \rbrack  \right)_f + \left(\; \frac{\mu A} {d} \right)_f \right]}_{A^{v_i}_{NB}} v^{n,m}_{i,NB}
=\\
-\sum_{f(\Delta V_P)}\left(p^{n,m-1}_{f}A_{i}\right)+
\rho_P \Delta V_P \left( g_i+ \frac{v^{n-1}_{i}}{\Delta t}
\right)_P +S_{v_i} \;,
\end{split}
\end{equation}
were $\dot {m}_f = (\rho v_i A_i)_f $ refers to the mass flux and $S_{v_i}$ to the term including deferred-corrections.

\subsection{Pressure correction equations}
\label{DiscrPress}
Here, the pressure correction equation is given by 
\begin{equation}
\sum_{f(\Delta V_P)} v^{m}_{f,i}A_{f,i}-\sum_{f(\Delta V_P)} \left(\frac{A_{i} d_i}{A_{P}^{v_i}}\right)_{f} \left(\frac{\partial p^{'}}{\partial x_i} \right)_{f} A_{f}=0\;,
\label{presscorr}
\end{equation}
where $p^{'}$ denotes the pressure correction and in the first stage the partial derivative of the pressure is discretized as 
\begin{equation}
\left(\frac{\partial p^{'}}{\partial x_i}\right)_{f}=\left(\frac{p^{'}_{NB}-p^{'}_{P}}{d}\right)\;.
\label{pressdervdiscr}
\end{equation}
leading to
\begin{equation}
\sum_{\Delta V_P} \Biggl[ \underbrace{\left(\frac{A_{i} d_i}{A_{P}^{v_i} d}\right)_{f}}_{A^{p^{'}}_P} p^{'}_{P}+\sum_{NB} \underbrace{-\left(\frac{A_{i} d_i}{A_{P}^{v_i} d}\right)_{f}}_{A^{p^{'}}_{NB}} p^{'}_{NB}=-\dot{v}^m_{\Delta V_P} \Biggr]\;.
\label{presseqalgsys}
\end{equation}
Afterward, the pressure $p^{m-1}$ is updated to $p^{m}$ with the obtained pressure correction $p^{'}$ ($p^{m}=p^{m-1}+p^{'}$). The velocity and fluxes are corrected via 
\begin{equation}
v^{m*}_{i,P}=v^{m}_{i,P}+\left(-\frac{\Delta V_P}{A_P^{v_{i,P}}}\right) \left( \frac{\partial p^{'}}{\partial x_i}\right)_P
\label{corrVel}
\end{equation}
and
\begin{equation}
\dot{v}^{m*}_f=\dot{v}^{m}_f+A^{p^{'}}_{NB} \left(p^{'}_{NB}-p^{'}_{P}\right)
\label{corrFlux}
\end{equation}
respectively. A Rhie-Chow \cite{RhieChow} interpolation of the fluxes is used to avoid pressure oscillations occurring for the collocated variable arrangement.\\
In the second stage of the pressure correction algorithm, the  non-orthogonal terms are included in the partial derivative of the pressure
\begin{equation}
\left(\frac{\partial p^{'}}{\partial x_i}\right)_{f}=\left(\frac{p^{'}_{NB}-p^{'}_{P}}{d}\right)+\left(\frac{\partial p^{'}}{\partial x_i} \left(\frac{A_i}{A}-\frac{d_i}{d}\right)\right)_f\;.
\label{press2stage}
\end{equation}
After solving the second equation system for the pressure correction, the pressure, fluxes and velocities are corrected again.

\subsection{Discretized Soil Mixture Fraction Equation}
A first-order implicit time integration scheme and an implicit upwind-difference scheme for the approximation of the convective term is applied to the soil mixture fraction equation
\begin{equation}
\begin{split} c^{n,m}_{P} \left[
\frac{\Delta V_P}{\Delta t} 
+\sum_{f(\Delta V_P)} A^{c}_{NB} \right] 
-\sum_{f(\Delta V_P)}
\underbrace{\left(max \left \lbrack - (\dot{m}/\rho)^{n,m-1},0 \right \rbrack \right)_f}_{A^{c}_{NB}} c^{n,m}_{NB}
=\frac{\Delta V_P }{\Delta t}c^{n-1}_{P}+S_{c} \, , 
\end{split}
\end{equation}
where $S_{c}$ hosts the deferred correction terms of the compressive approximation. The mass flux $\dot{m}$ is calculated from the corrected flux $\dot{v}^{m*}$.

\subsection{Discretized Displacement Equation}
The Eulerian displacement equation \eqref{EQDisplFV} is approximated applying a first-order implicit time integration scheme and an implicit upwind-difference scheme for the convective term. The discrete equation is then given by
\begin{equation}
\begin{split}
u^{n,m}_{i,P} \left[\frac{\Delta V_P}{\Delta t} +\sum_{f(\Delta V_P)} A^{u}_{NB} \right] -\sum_{f(\Delta V_P)}\underbrace{\left(max \left \lbrack - \dot{v}^{S ,n,m-1},0 \right \rbrack\right)_f }_{A^{u}_{NB}} u^{n,m}_{i,NB}=\\
\frac{\Delta V_P }{\Delta t}u^{n-1}_{i,P}+ c_{P} v^{n,m-1}_{i,P} \Delta V_P+S_u\;,
\end{split}
\end{equation}
where $\dot {v}^S_f = (c v_i A_i)_f $ refers to the flux, and $S_u$ again includes explicit terms which arise from different deferred correction contributions, e.g., higher-order convection, non-orthogonality, and interpolation corrections.

\end{document}